\documentclass[10pt,letterpaper]{article} 

\usepackage{titlesec,float,makeidx,multirow,multicol,graphicx,hyperref,amsmath,amssymb,fancyhdr,changepage,etoolbox,times,enumitem,textcomp,indentfirst,subcaption,caption,setspace}

\usepackage[dvipsnames,svgnames,table]{xcolor}
\usepackage{txfonts}
\usepackage{mathdots}
\usepackage[paperwidth=8.5in,paperheight=11in,top=1in,right=0.5in,bottom=1in,left=0.5in]{geometry}
\usepackage[utf8]{inputenc}
\usepackage[sort&compress,numbers]{natbib} 
\fancypagestyle{firstpage}{
\fancyhead[L,C,R]{}
\fancyfoot[R]{}
\fancyfoot[C]{\thepage}
\fancyfoot[L]{}

}

\RequirePackage{fix-cm}
\newcommand{\size}[2]{{\fontsize{#1}{0}\selectfont#2}}

\titleformat{\section}
  {\normalfont\fontfamily{phv}\bfseries}{\thesection}{10pt}{\MakeUppercase}
\renewcommand*{\thesection}{\fontfamily{phv}\selectfont\textbf{\arabic{section}.}}
\titlespacing{\section}{0pt}{10pt}{0pt}
\titleformat{\subsection}
  {\normalfont\fontfamily{phv}\bfseries}{\thesubsection}{3pt}{}
\renewcommand*{\thesubsection}{\fontfamily{phv}\selectfont\textbf{\arabic{section}.\arabic{subsection}}}
\titlespacing{\subsection}{0pt}{10pt}{0pt}

\DeclareCaptionFormat{upper}{#1#2\uppercase{#3}\par}
\newcommand{\startsquarepar}{%
    \par\begingroup \parfillskip 0pt \relax}
\newcommand{\stopsquarepar}{%
    \par\endgroup}

\hypersetup{
	colorlinks=true,
	linkcolor=black,
	filecolor=blue,      
	urlcolor=blue,
	citecolor=black,
}
\usepackage{nomencl}
\usepackage{bm}
\usepackage{xstring}
\usepackage{xpatch}
\patchcmd{\thenomenclature}
  {\leftmargin\labelwidth}
  {\leftmargin\labelwidth\itemindent 0.25in }
  {}{}

\makenomenclature
\renewcommand{\nomname}{NOMENCLATURE}
\begin{document}
\newgeometry{top=0.5in,right=0.5in,bottom=1in,left=0.5in}
\begin{flushright}
\fontfamily{phv}\selectfont{
\textbf{Conference Paper}\\
\textbf{2022}\\
\textbf{ }\\
\textbf{ }\\[45pt]
\textbf{\size{18}{ }}\\[38pt]
}
\end{flushright}
\begin{center}
    \fontfamily{phv}\selectfont{\size{11}{\textbf{Development of a Finite Element Solver Including a Level-Set Method for Modeling Hydrokinetic Turbines\\[20pt]}}}
\end{center}


    \begin{flushright}
        \fontfamily{phv}\selectfont{\textbf{Ahmed A. Hamada$^{1}$, Mirjam Fürth$^{1}$} \newline
        $^1$Department of Ocean Engineering, Texas A$\&$M University, College Station, TX, 77843, USA \newline }
    \end{flushright}
    
\begin{multicols*}{2}
\section*{Abstract}
Hydrokinetic flapping foil turbines in swing-arm mode have gained considerable interest in recent years because of their enhanced capability to extract power, and improved efficiency compared to foils in simple mode. The performance of foil turbines is closely linked to the development and separation of the Leading-Edge Vortex (LEV). To accurately model the formation and the separation of the LEV on flapping foils, a purpose-built 2D numerical model was developed. The model is based on the weighted residual Finite Element Method (FEM); this is combined with an interface capturing technique, Level-Set Method (LSM), which was used to create a reliable and high-quality numerical solver suitable for hydrodynamic investigations. The solver was validated against well-known static and dynamic benchmark problems. The effect of the mesh density was analyzed and discussed. This paper further covers an initial investigation of the hydrodynamics of flapping foil in swing-arm mode, by studying the structure of the vortex around a NACA0012 foil. The presented method helps to provide a better understanding of the relation between the Leading-Edge Vortex creation, growth, and separation over the flapping foil in swing-arm mode and the extracted power from a hydrokinetic turbine.

Keywords: Finite Element Method; Level-Set Method; current/tidal energy harvesting; hydrokinetic foil turbine; swing-arm mode.
\mbox{}
\nomenclature[A]{$\textbf{V}$}{Flow field velocity vector}
\nomenclature[A]{$U$}{Free-stream velocity}
\nomenclature[A]{$Re$}{Reynolds number}
\nomenclature[A]{$p$}{Flow field pressure}
\nomenclature[A]{$t^*$}{Dimensionless time w.r.t the flapping period}
\nomenclature[A]{$T$}{Flapping period}
\nomenclature[A]{$k$}{Reduced frequency}
\nomenclature[A]{$c$}{Chord of flapping foil}
\nomenclature[B]{$\omega$}{Angular frequency of flapping foil}
\nomenclature[A]{$k$}{Reduced frequency}
\nomenclature[A]{$S$}{Swing factor}
\nomenclature[A]{$h_o$}{Heave amplitude}
\nomenclature[B]{$\alpha$}{Free-Stream angle of attack}
\nomenclature[A]{$C_P(t)$}{Instantaneous power coefficient}
\nomenclature[A]{$C_T(t)$}{Instantaneous thrust coefficient}
\nomenclature[B]{$\eta$}{Generating power efficiency}
\nomenclature[A]{FEM}{Finite Element Method}
\nomenclature[A]{LSM}{Level-Set Method}
\nomenclature[A]{$w_i(.)$}{Weight function in FEM}
\nomenclature[A]{$N_i$}{Shape function within the elements}
\nomenclature[B]{$\tau$}{Local mesh stabilization parameter}
\nomenclature[A]{$V_{LS}$}{Convective velocity vector of LS function}
\nomenclature[B]{$\phi$}{Level-Set function}
\printnomenclature[0.68in]
\section{Introduction}
\startsquarepar \indent The demand for clean electricity from alternative power sources is rising as more consumers value sustainability \cite{armaroli2011towards}. Ocean currents have great potential to generate reliable clean power \cite{freris2008renewable}, and a range of market-ready hydrokinetic turbines have been developed over the last twenty years. For example, the Oscillating Marine Current Energy Converter (“Stingray”) \cite{Stingray2002Phase1, Stingray2003Phase2, Stingray2005Phase3}, and the 1kW Oscillating Hydrofoil Energy Harvesting System \cite{cardona2016field}. Furthermore, current energy has less environmental impact than conventional hydroelectric facilities, such as dams, diversions, and pumped storage \cite{lago2010advances}. 

\begin{figure}[H]
	\centering
	\includegraphics[width=\linewidth]{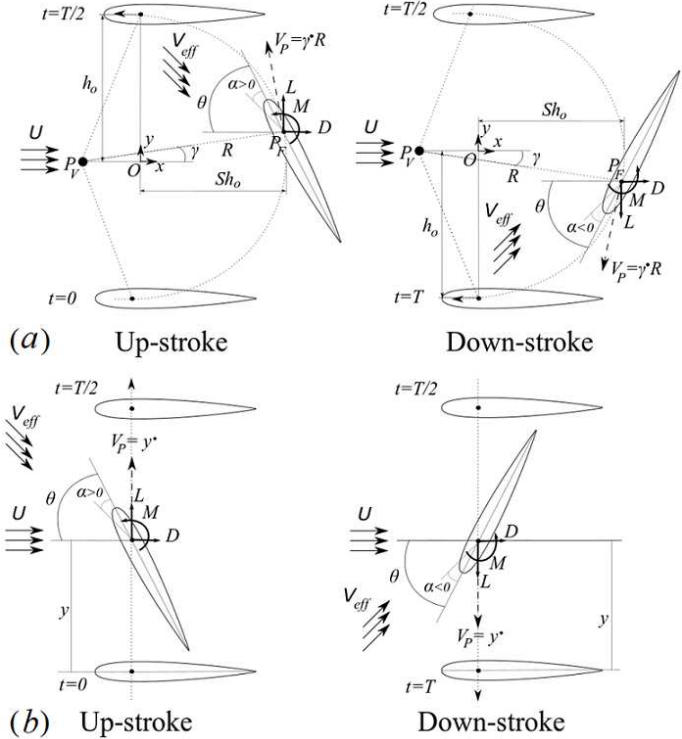}
	\caption{A flapping foil system with different motion patterns: (a) Swing-arm power extraction mode, and (b) Simple power extraction mode. In each mode, the associated angle of attack $\alpha$, and force directions during the flapping cycle are shown}
	\label{fig:flapping_foil}
\end{figure}

Flapping foils allow for energy extraction in free surface waves, open channels, and uniform currents \cite{fernandes2013marine}. Wu et al. \cite{wu1972extraction} introduced the concept of generating energy using an oscillating wing in free-surface waves in the 1970s. They showed that a flapping foil generates power by performing two main motions; heave and pitch. Oscillating foil turbines generate a leading-edge vortex (LEV), which leads to higher forces. As a result, they create higher lift forces than conventional rotary turbines in which the flow remains attached to the turbine blades \cite{young2014review}. Furthermore, by adopting a swing-arm mode, in which the heave motion is performed by an arm rotation, the efficiency can further be improved compared to simple mode, in which the heave motion is performed simply up and down \cite{wu2015power}. The swing-arm mode flapping foil (a) is shown in Figure \ref{fig:flapping_foil} in comparison with a simple mode foil (b). Karbasian et al. \cite{karbasian2016power} showed that the swing-arm mode changes the LEV's\stopsquarepar
\end{multicols*}
\restoregeometry
\clearpage
\begin{multicols*}{2} 
\noindent creation, growth, separation, and delay (in time). Furthermore, more power can be extracted by selecting the optimum value of the swing factor. Numerical simulations allow for deep investigations of the formation LEV around a flapping foil, which may require very advanced motion capture equipment and other experimental setups \cite{boudreau2018experimental,duarte2019experimental}. Thus, numerical methods allow faster evaluation of new designs compared to experiments.
 
In this paper, the water flow around the flapping foil is modeled with the incompressible Navier–Stokes (NS) equations. To numerically solve the NS equations, a discretization technique has to be used, such as Finite Difference Method (FDM), Finite Element Method (FEM), or Finite Volume Method (FVM). FEM works on the weak (integral) form of the partial differential equations (PDE), with a lower derivative order compared to the strong (differential) form. The order of the approximated solution in the weak form is half of that in the strong form \cite{martin1973introduction}. This makes FEM more powerful than FDM, which discretizes the strong form directly. FEM has an advantage compared to other numerical methods, such as FDM, and FVM, in the application of boundary conditions \cite{lewis2004fundamentals}. Because FEM applies only physical boundary conditions directly for the nodal values of the field variables, there is no need for numerical boundary conditions. In addition, the natural and mixed boundary conditions are automatically inherent in the weak form of the PDE \cite{donea2003finite}. Further, FEM is better at handling unstructured meshes \cite{argyris1979finite}.

There are two main types of meshes, moving and fixed \cite{mastin1985numerical}. In moving meshes, the coordinate nodes follow the moving boundaries; i.e. the mesh is reprocessed at each time step which consumes computational power \cite{liseikin1999grid}. Further, the moving meshes are more likely to lose their quality due to the element inversion after several time steps. This requires a re-meshing that increases computational time \cite{carey1997computational}. The coordinate nodes do not change in fixed meshes, making them preferred \cite{edelsbrunner2001geometry}. Instead, the boundaries are defined with an interface capture technique, such as Volume of Fluid \cite{hirt1981volume,ashgriz1991flair}, Marker and Cell technique \cite{harlow1965numerical}, Level-Set technique \cite{osher2006level}, and Diffuse interface technique  \cite{anderson1998diffuse}. During the last two decades, Level-Set Method (LSM) has experienced increased usage in hydrodynamics due to its simplicity in capturing the moving boundaries \cite{di2007application,ali2021stress}, and the accuracy of obtained results \cite{sharma2015level, elbadry2021active}. Kandasamy et al. \cite{kandasamy2005vortical} showed the best prediction of the wave-induced separation and the wave profile on a surface piercing NACA0024 foil by using the LSM with the Unsteady Reynolds Averaged Navier Stokes (URANS) and Detached Eddy Simulation (DES). The reason is that the LSM is suitable to simulate the shear layer instability problems, such as turbulent/splashing free-surface, unlike the surface tracking method \cite{brocchini2002free}. Further, Chung \cite{chung2016propulsive} Used the LSM to capture the free surface near a flapping flat plate. Moreover, Bergmann and Iollo \cite{bergmann2016bioinspired} showed the applicability of using the LSM in simulating bio-inspired swimmers, such as fish.

Building a purpose-built code is very beneficial because it offers the controllability and flexibility to integrate the powerful solvers, FEM and LSM. This allows for simulations of incompressible flow while capturing the flapping motion of the foil. Thus, a better understanding of how LEV formation and separation affect the performance of swing arm flapping foils and their relation to the energy harvested can be provided by studying the hydrodynamics and vortex structure around the flapping hydrokinetic energy converter. 
\section{Computational Model}
 Foil in simple mode (Figure \ref{fig:flapping_foil} (b)) moves only in the $y$ direction. However, in swing-arm mode (Figure \ref{fig:flapping_foil} (a)), the foil rotates around the base pivoting point, $P_B$, and the radius of rotation is the length of the arm, $R$.The swing-arm mode increases the movement in $x$ and $y$ directions. Up-stroke and down-stroke define the motion in the $y$ direction, whereas forward and backward motions are in the $x$ direction. The hydrofoil moves on an arc during the cycle. The origin, $O$, is chosen to be at the middle of the line joining the extreme positions. The maximum distance from $O$ vertically is the heave amplitude, $h_o$, and the maximum horizontal distance is $R-Sh_o$, where $S$ is the swing factor. In addition, the foil performs a second rotational motion around its pivoting point, $P_F$, during its cycle. Hence, the hydrodynamic and flywheel inertia forces enable the foil to rotate the connected swing-arm, which rotates the generator shaft at the base pivoting point, $P_V$, to generate power. 
\subsection{Governing Equations}
The influence of using swing-arm mode on the LEV, created over an oscillating foil was modeled using 2D Navier-Stokes equations for unsteady incompressible laminar flow \cite{wang2010numerical}. The dimensionless governing equations for the flow around an oscillating foil may be expressed as:
\begin{equation} \label{eq:Continuity}
\nabla \cdot \textbf{V} = 0
\end{equation}
\begin{equation} \label{eq:Momentum}
\frac{k}{2\pi} \frac{\partial \textbf{V} }{\partial t^*}+\left(\textbf{V} \cdot \nabla \right) \textbf{V}= - \nabla p+\frac{1}{Re} \nabla^2 \textbf{V}
\end{equation}
where $\textbf{V}$ is the dimensionless velocity vector of the flow in 2D Cartesian coordinates with respect to the free-stream velocity, $U$, $t^*$ is the dimensionless time with respect to the flapping period, $T$, $p$ is the dimensionless pressure of the flow with respect to the dynamic pressure, $\rho U^2$, ($\rho$ is the density of flow), $Re$ is the free-stream Reynolds number, and $k$ is the non-dimensional reduced frequency parameter, which results from the non-dimensionalization of time, and is defined as:
\begin{equation} 
k=\frac{\omega c}{U} 
\end{equation}
where $c$ is the foil's chord (the characteristic length), and $\omega=2 \pi f$ is the angular frequency of the oscillating foil.
\subsection{Incompressibility Constraint}
The steady-state solution is obtained using an iterative process over the whole domain to satisfy the incompressible continuity equation \cite{kwak2005computational}. In this study, the pressure stabilization technique is used to restore the coupling between continuity and momentum equations \cite{langtangen2002numerical,elhanafy2017pressure}. This technique modifies the continuity equation by adding a Laplacian term as follows:
\begin{equation} \label{eq:incomp_constraint}
\nabla \cdot \textbf{V} = \epsilon \nabla^2 p
\end{equation}
where $\epsilon$ is the pressure dissipation parameter; its order of magnitude is the same as the dimensionless time step, $\Delta t^*$.
\subsection{Kinematics}
\begin{figure}[H]
	\centering
	\includegraphics[width=\linewidth]{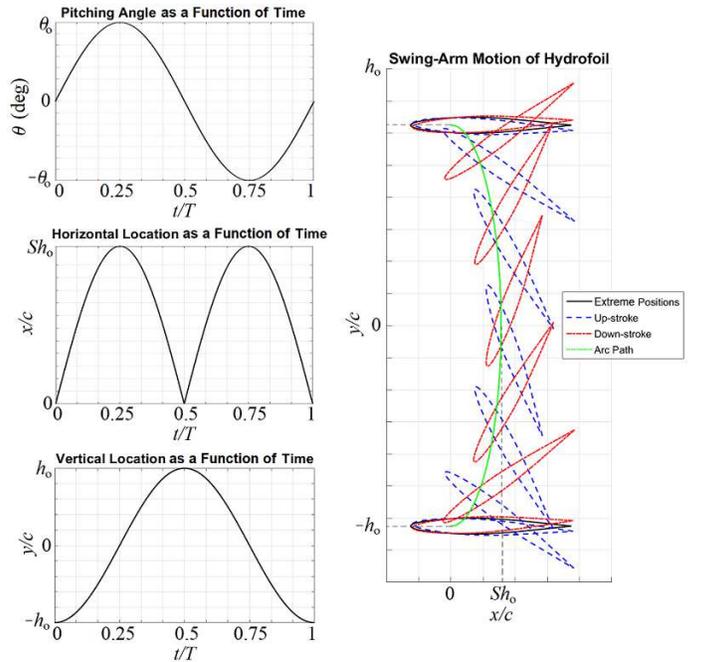}
	\caption{Kinematics (left) and foil motion (right) of flapping foil in Swing-arm mode. The right figure shows the foil position and orientation during the up-stroke and down-stroke. The arc path indicates the location of the foil's pivoting point, $P_F$. The extreme positions show the foil at maximum heave amplitude.}
	\label{fig:Kin}
\end{figure}

As shown in Figure \ref{fig:Kin}, the flapping foil in swing-arm mode with a harmonic sinusoidal function experiences simultaneous pitching motion, $\theta (t)$, and plunge motion in Cartesian coordinates, $x(t)$ and $y(t)$, which are respectively defined as:
\begin{subequations}
	\begin{equation} \label{eq:Pitching_M}
	\theta (t) = \theta_o \sin(\omega t)
	\end{equation}
	\begin{equation} \label{eq:x_M}
	x(t)=Sh_o \left| \sin \left(\omega t\right) \right|
	\end{equation}
	\begin{equation} \label{eq:y_M}
	y(t)=-h_o \sin \left(\omega t+\psi\right) 
	\end{equation}
\end{subequations}

where $\theta_o$ is the pitch amplitude, $\psi$ is the phase difference angle between pitching and plunging motions, $h_o$ is the heave amplitude, and $S$ is the swing factor whose value is in the range of $\left(0,1\right]$ for the swing-arm mode. When $S$ is zero, the motion reverts to a simple flapping foil. The pitching velocity, $\Omega(t)$, and the upstream velocity components, $V_x(t)$ and $V_y(t)$, of a flapping foil due to the swing-arm mode are respectively expressed as:

\begin{subequations}
	\begin{equation} \label{eq:Vp}
	\Omega(t)=\frac{d\theta(t)}{dt}=\theta_o \omega \cos\left(\omega t\right) 
	\end{equation}
	\begin{equation}\label{eq:Vx}
	\begin{split}
	V_x&(t)=\frac{dx(t)}{dt} \\ &= \left\lbrace
	\begin{matrix}
	Sh_o \omega \cos \left(\omega t\right) & 0\leq \frac{t}{T}\leq \frac{1}{2}\\-Sh_o \omega \cos \left(\omega t\right) & \frac{1}{2}<\frac{t}{T}\leq 1
	\end{matrix}
	\right\rbrace
	\end{split}
	\end{equation}
	\begin{equation} \label{eq:Vy}
	V_y(t)=\frac{dy(t)}{dt}=-h_o \omega \cos \left(\omega t+\psi\right) 
	\end{equation}
\end{subequations}
		
The plunge motion of the hydrofoil introduces an induced angle of attack. Thus, the effective angle of attack, $\alpha_{eff}$, and the effective upstream velocity, $V_{eff}$, are:
\begin{equation} \label{eq:alpha_eff}
\alpha_{eff}(t)=\theta(t)+\arctan\left(\frac{V_y}{U+V_x}\right) 
\end{equation}
\begin{equation} \label{eq:V_eff}
V_{eff}(t)=\sqrt{\left(U+V_x\right)^2+\left(V_y\right)^2}
\end{equation}
		
The maximum effective angle of attack, $\alpha_{max,eff}$, which is approximately the modulus of the angle of attack at quarter-period, highly affects the generation of peak forces over the hydrofoil and occurrence of dynamic-stall \cite{simpson2008experiments}. The maximum effective upstream velocity, $V_{max,eff}$, also happens at the same time as the maximum effective angle of attack, $\alpha_{max,eff}$, and they are defined as: 
\begin{equation} \label{eq:alpha_eff_max}
\begin{split}
\alpha_{max,eff} =\left|\alpha_{eff}\left(\frac{T}{4}\right) \right|=\left| \theta_o-\arctan\left(\frac{h_o \omega}{U}\right) \right|
\end{split}
\end{equation}
\begin{equation} \label{eq:V_eff_max}
\begin{split}
V_{max,eff}=V_{eff}\left(\frac{T}{4}\right)=\sqrt{\left(U\right)^2+\left(h_o \omega\right)^2}
\end{split}
\end{equation}

The feathering parameter, $\chi$, that indicates power extraction or propulsion, is defined as:
\begin{equation} \label{eq:Feathering}
\chi=\frac{\theta_o}{\arctan\left(\frac{h_o \omega}{U}\right)}
\end{equation}

The necessary but not sufficient condition for power extraction mode is $\chi>1$, i.e., $\alpha_{eff}\left(\frac{T}{4}\right)>0$. However, the condition for propulsion mode is $\chi<1$, i.e., $\alpha_{eff}\left(\frac{T}{4}\right)<0$.
\subsection{Extracted Power And Efficiency}
For power extraction mode, $\chi>1$, the instantaneous power and thrust coefficients, $C_P$ and $C_T$, depend on the contributions from both the rotational plunge and pitching motions and are respectively defined as:
\begin{equation} \label{eq:Power_Coeff}
C_P(t) =\frac{F_y V_y+F_x V_x+M \Omega}{0.5 \rho c U^3}
\end{equation}
\begin{equation} \label{eq:Thrust_Coeff}
C_T(t) =\frac{F_x V_x}{0.5 \rho c U^2}
\end{equation}
where $F_x$, $F_y$, and $M$ are the magnitudes of the instantaneous forces and pitch moment at the foil pivoting point, $P_F$, in the Cartesian coordinates; as a result, the mean of either power or thrust coefficients, $\overline{C}_P$ or $\overline{C}_T$, can be calculated over one cycle.
Finally, the efficiency, $\eta$, which indicates the capability of the oscillating foil system to generate power from the flow, is defined as:
\begin{equation} \label{eq:effi}
\eta =\frac{\overline{P}_{extracted}}{P_{available}}=\frac{c}{2h_o} \overline{C}_P
\end{equation}
\section{Numerical solver}
\subsection{Finite Element Formulation}
The numerical model is based on the weighted residual FEM. FEM redefines the variables by separating their spatial and temporal dependence using the shape functions, $N_i(.)$ \cite{hughes2012finite}. The weighted residual approach minimizes the weighted error of the solution for the weak form of the PDE \cite{connor2013finite}. Furthermore, the residual approach varies by changing the value of the weighting parameter \cite{segerlind1976applied}. Hamada et al. \cite{Hamada2020Variants} showed that the up-winding Galerkin$/$Least-Square (GLS), which was introduced by Hughes et al. \cite{hughes1989new}, is the best Weighting approach to solve second-order differential equations with convection terms, such as the momentum Navier-Stokes equations. This is because of its stability and accurate diffusivity compared to the  Least-Square, Standard Galerkin, Collocation, Galerkin$/$Least-Squares, Collocation$/$Galerkin, Collocation$/$Least-Squares, and Collocation$/$Galerkin$/$Least-Squares. On the other hand, there are not any convective terms in the modified continuity equation and its type is elliptic. Therefore, the Standard Galerkin (SG) formulation is enough to solve the modified continuity equation \cite{thomee1984galerkin}. A forward first-order explicit scheme \cite{strikwerda2004finite} was chosen to conduct the time marching term. The weight function, $w_i(.)$, of SG and GLS are:

\begin{equation}
\label{eq:weight_fn}
w_i(.)=
\left\lbrace
\begin{matrix}
N_i(.) & SG \\ N_i(.)+\tau L(N_i(.)) & GLS
\end{matrix}
\right\rbrace
\end{equation}
where $N_i(.)$ is the shape functions within the elements, $L(N_i(.))$ is the Least-Square operator over the shape functions which is constructed from the residual of the governing equation, and $\tau$ is the local mesh stabilization parameter, which is defined as:  
\begin{equation}
\label{eq:tau}
\tau=\frac{1}{\sqrt{\left(\frac{k}{\pi \Delta t^*}\right)^2+\left(\frac{2V_e}{L_e}\right)^2+\left(\frac{4}{L_e^2 Re}\right)^2}}
\end{equation}
where $V_e$ and $L_e$ are the total velocity and characteristic length of each element, respectively. 
	
Integration by parts was performed for the Laplacian term to reduce the order of partial differential equations which enables the usage of bi-linear shape functions and allows the use of the Neumann boundary condition for the pressure if needed \cite{hughes2012finite}. Moreover, the mass lumping technique is used for the unsteady term matrix to reduce computational time \cite{bicanic1991mass,wendland2005numerical}. The weak forms of the continuity and momentum equations in indicial notation are:

\begin{equation}
 \label{Cont_WeakForm}
 \frac{-1}{\epsilon} \iint N_i \nabla \cdot \textbf{V} dA =\iint \nabla N_i \cdot \nabla p dA-\oint N_i \frac{\partial p}{\partial n} dS 
\end{equation}
\begin{equation}
 \begin{split}
 \label{eq:Momen_Weak}
 \frac{k}{2\pi}& \iint N_i \frac{\Delta \textbf{V}}{\Delta t^*} dA +\iint \widetilde{N}_i \left(\textbf{V}_e \cdot \nabla \right) \textbf{V} dA  =-\iint N_i \nabla p dA \\&- \frac{1}{Re} \left( \iint \left(\nabla N_i \cdot \nabla \right) \textbf{V} dA + \oint N_i \frac{\partial \textbf{V}}{\partial n} dS \right) 
 \end{split}
\end{equation}
\subsection{Level-Set Method}
\begin{figure}[H]
   \centering
   \includegraphics[width=\linewidth]{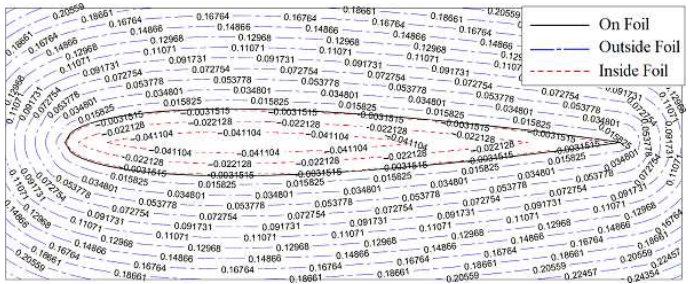}
   \caption{Contour levels of foil using LSM; the levels inside the foil are negative, the levels outside the foil are positive, and the level on the foil has a zero value.}
   \label{fig:LSM_Contour}
\end{figure}

LSM is an interface capturing method which defines the nodes of the body in the fixed meshes using an implicit function, $\phi$ \cite{osher2006level}. The function, $\phi$ is constant along the contour levels around the foil's surface, see Equation (\ref{eq:LSMeq}). The value of the LSM function classifies the domain into three regions; inside, on, and outside the foil surface, as seen in Figure \ref{fig:LSM_Contour}. Thus, the boundaries of the foil will be determined with non-positive contour levels. Further, LSM prevents element inversion and maintains the quality of the mesh, unlike the dynamic meshes which require re-generation with time \cite{osher2001level}. The boundary of the foil is located dynamically by a Signed-Distance Function (SDF) at each time step \cite{osher2003signed}, see Figure \ref{fig:Mesh_LSM_foil} right.
\begin{equation} \label{eq:LSMeq}
\frac{D \phi}{\partial t}=\frac{\partial \phi}{\partial t}+ \boldsymbol{V_{LS}} \cdot \nabla \phi = 0
\end{equation}
where $\boldsymbol{V_{LS}}$ is the convective velocity vector of the level set function, $\phi$.
\subsection{Mesh Design}
\begin{figure}[H]
	\centering
	\includegraphics[width=\linewidth]{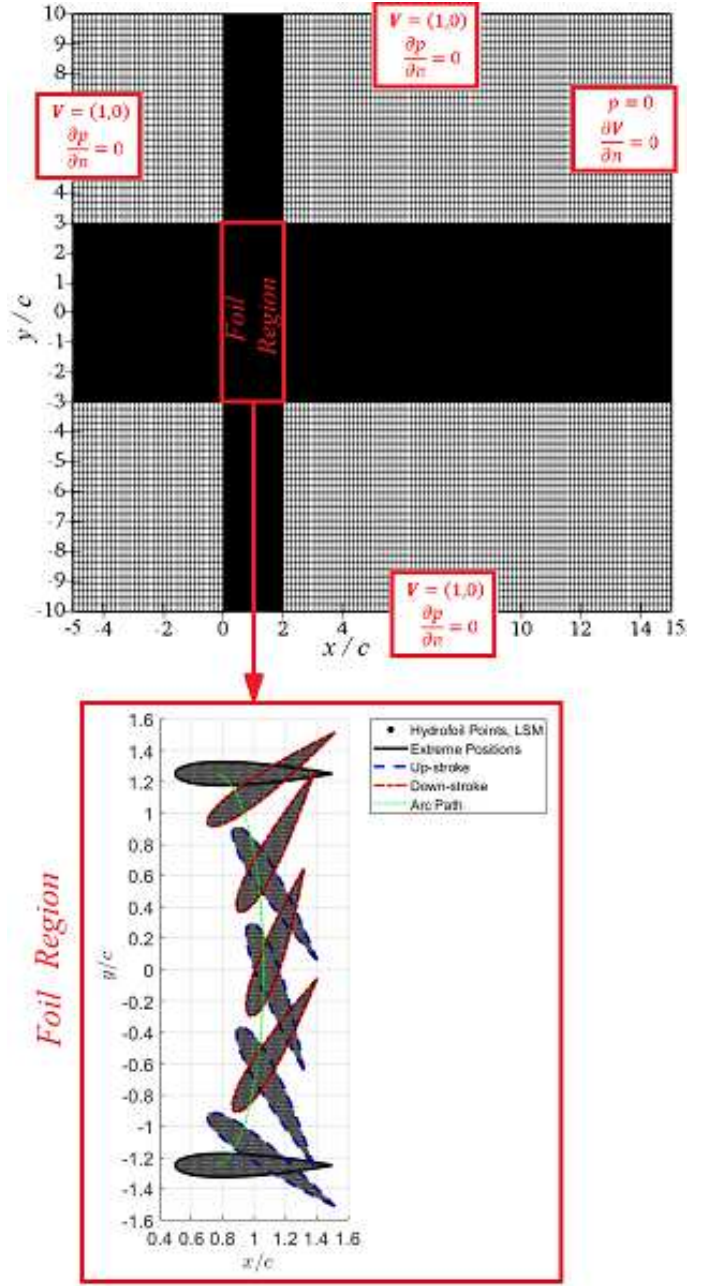}
	\caption{Characteristics of the computational domain (top), showing the mesh, the clustered foil region, and the boundary condition. The flapping foil (bottom) at certain time steps using LSM, indicating the refinement of the mesh at the foil region and the quality of the LSM.}
	\label{fig:Mesh_LSM_foil}
\end{figure}

\begin{figure}[H]
	\centering
	\includegraphics[width=\linewidth]{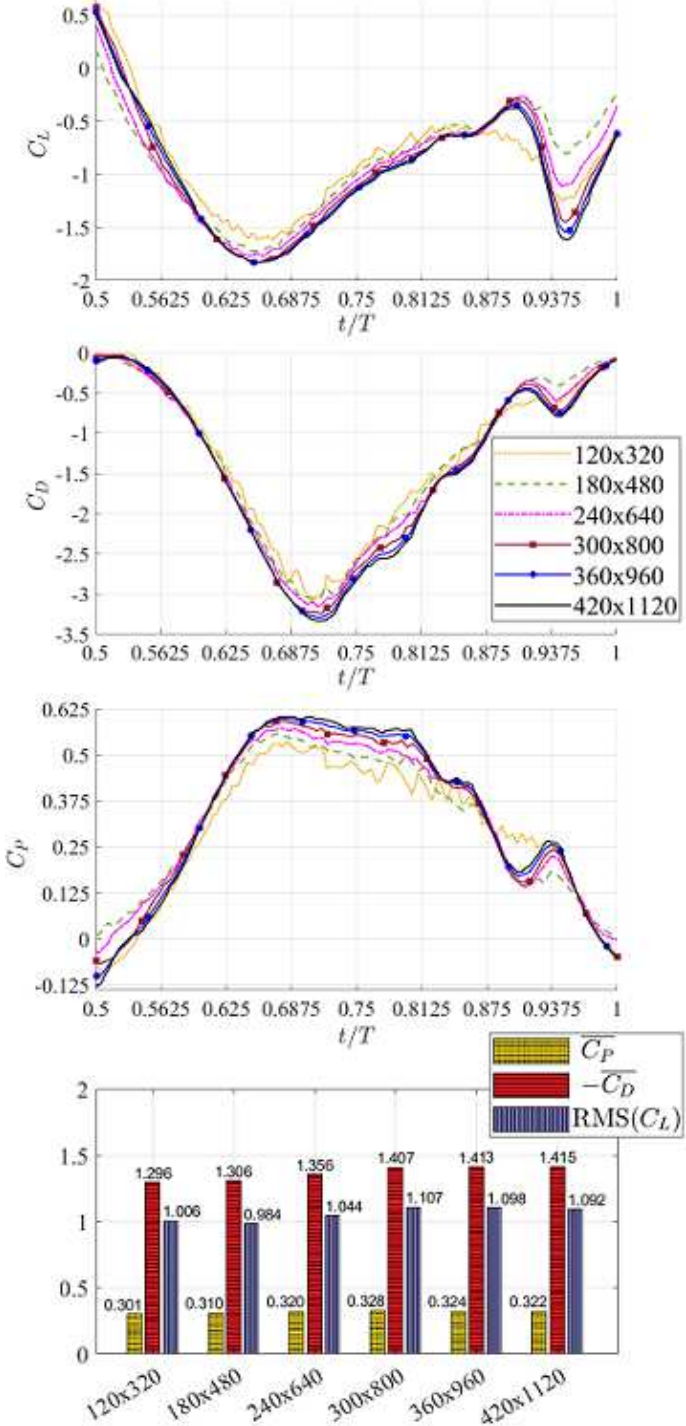}
	\caption{Independent mesh test: the lift, drag, and power coefficients for oscillating NACA0012 foil in swing-arm mode using different grid sizes at heave amplitude of $1.25$, pitch amplitude of $70^o$, swing factor of $0.25$, non-dimensional reduced frequency of $0.08$, and Reynolds number of $1\times10^5$.}
	\label{fig:MeshInd1}
\end{figure}

The mesh (Figure \ref{fig:Mesh_LSM_foil} left) is a structured square shape, does not include the foil, and has a length of $20 c$, where $c$ is the chord length. Mesh independence was checked to avoid grid dependent results. Six meshes, with a number of elements of $120 \times 320$, $180 \times 480$, $240 \times 640$, $300 \times 800$, $360 \times 960$, and $420 \times 1120$ were selected to perform the mesh independence, by simulating the oscillating NACA0012 foil in swing-arm mode with $Re=1\times10^5$, $k=0.08$, and $S=0.25$. Figure \ref{fig:MeshInd1} shows the instantaneous lift, drag, and power coefficients for the considered test during a cycle. The results of the meshes $300 \times 800$, $360 \times 960$, and $420 \times 1120$ are very similar, which means that the model becomes mesh independent, see Table \ref{table:2}. Hence, the change in results, with finer mesh, can be neglected. Thus, a mesh of $360 \times 960$ was selected.

\begin{table}[H]
\centering
\caption{Independent Mesh test: Absolute percentage error of the mean of power coefficient, negative mean of drag coefficient, and root mean square of lift coefficient during one flapping cycle with respect to the mesh of $420 \times 1120$ elements for oscillating NACA0012 foil in swing-arm mode.}
\begin{tabular}{c c c c} 
 \hline 
 \multirow{2}{*}{Number of elements}&\multicolumn{3}{c}{ $|Error|\%$}\\
 & $\overline{C_P}$ & -$\overline{C_D}$ & $RMS\left(C_L\right)$\\ [0.5ex]
 \hline
 $120 \times 320$ & $6.7284$ & $8.4181$ & $7.8392$ \\ 
 $240 \times 640$ & $0.5818$ & $4.1344$ & $4.3438$\\
 $360 \times 960$ & $0.5466$ & $0.3605$ & $1.3376$\\ [1ex] 
 \hline
\end{tabular}
\label{table:2}
\end{table}

The area of the flapping foil region is $2c \times 6c$ and the number of elements within that region is $240 \times 900$. The distance from the first node to the foil surface is around $6.67 \times e^{-3} c$. The top and bottom are located $7c$ from the flapping foil region, while the inlet boundary is located $5c$ from the flapping foil region. All these three boundaries are specified with a horizontal unity free stream and a zero pressure gradient. The downstream boundary is $13 c$ from the foil region, with a condition of zero velocity gradient and single-point gauge pressure. The oscillating foil points have a no-slip condition and zero pressure gradient.
\section{Validation Study}
The numerical solver was validated using four well-known benchmark problems, stationary circular cylinder, stationary foil, oscillating circular cylinder, and flapping foil with simple sinusoidal motion. 
\subsection{Stationary Circular Cylinder}
\begin{figure}[H]
	\centering
	 \begin{subfigure}[H]{\linewidth}
		 \centering
		\includegraphics[width=\linewidth]{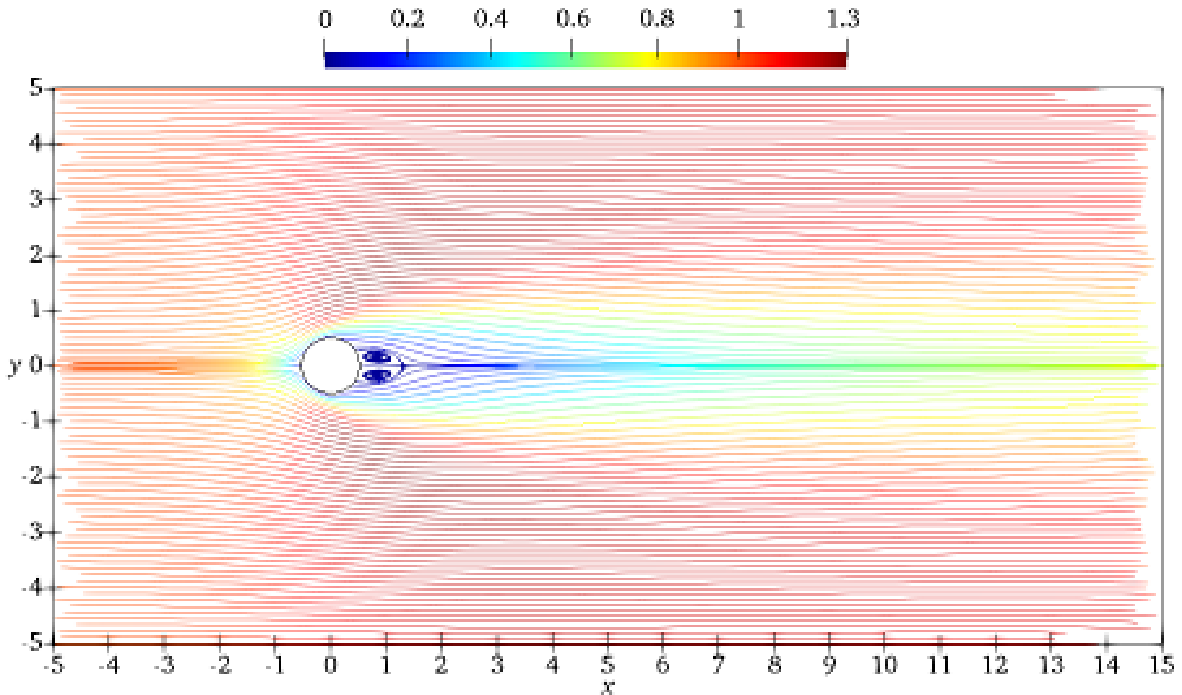}
		\caption{ at $Re$ = 20.}
		\label{fig:StCyRe20}
	\end{subfigure}\\
	\begin{subfigure}[H]{\linewidth}
		\centering
		\includegraphics[width=\linewidth]{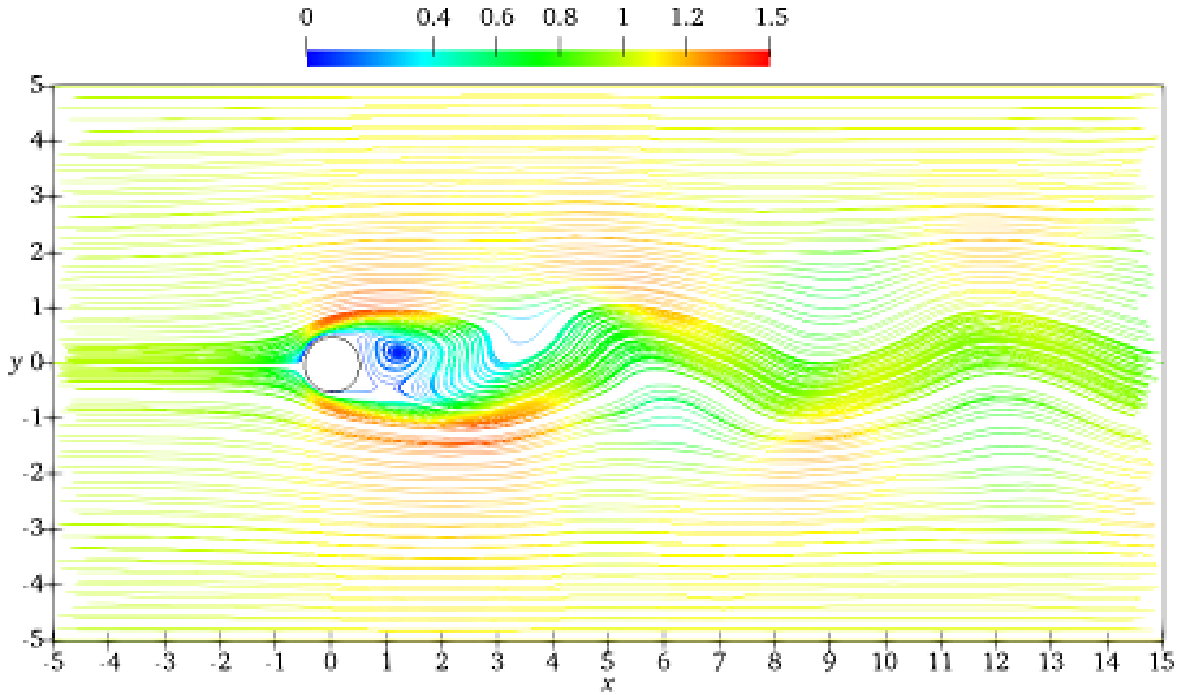}
		\caption{ at $Re$ = 300.}
		\label{fig:StCyRe300}
	\end{subfigure}
	\caption{Streamlines of flow around a stationary circular cylinder using FEM at different Reynolds numbers.}
    \label{fig:StCyDiffRe}
\end{figure}

The stationary circular cylinder problem is well-known and often used for studying the characteristics of the vortex shedding phenomenon, where the flow behavior is controlled by the Reynolds number \cite{fredsoe1997hydrodynamics}. As the Reynolds number increases, the flow changes from generating symmetric wakes to periodic vortex shedding behind the circular cylinder. Two simulations of flow past a stationary circular cylinder were performed to validate the FEM solver using a body-fitted mesh; two reference cases were chosen for the validation. Huang et al. \cite{huang2003research} showed that the vortex length to cylinder length ratio is about $0.991$, while the present solver obtained a vortex length to cylinder length ratio of $0.941$ at $Re=20$. Williamson \cite{williamson1989oblique} experimentally obtained a  Stroughal number, $St=\frac{f d}{U}$, where $f$ is the shedding frequency, $d$ is the diameter of the circular cylinder, and $U$ is the free-stream fluid velocity, of $0.202$ at a Reynolds number of $300$. The Stroughal number using the present solver was  $0.17$. The streamlines around the cylinder at the two Reynolds numbers are shown in Figure \ref{fig:StCyDiffRe}, showing the validity of the purpose-built FEM solver to simulate the problems of incompressible flows.
\subsection{Stationary Foil}
\begin{figure}[H]
	\centering
	\includegraphics[width=\linewidth]{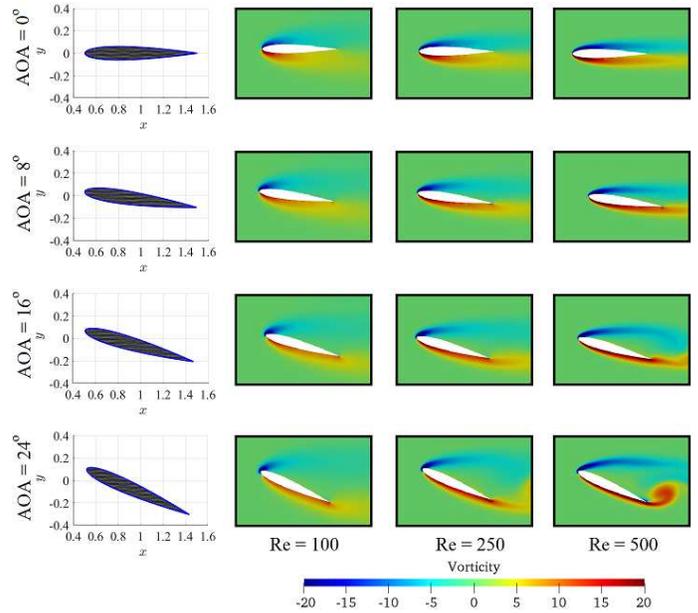}
	\caption{Stationary foils points (left) and vorticity contours (right) around the foils at different angles of attacks, $AOA=0^o$, $8^o$, $16^o$, $\&$ $24^o$, and different Reynolds numbers, $Re=100$, $250$, $\&$ $500$.}
	\label{fig:FlapFoil_Sta}
\end{figure}

\begin{figure}[H]
	\centering
	 \begin{subfigure}[H]{0.49\linewidth}
		 \centering
		\includegraphics[width=\linewidth]{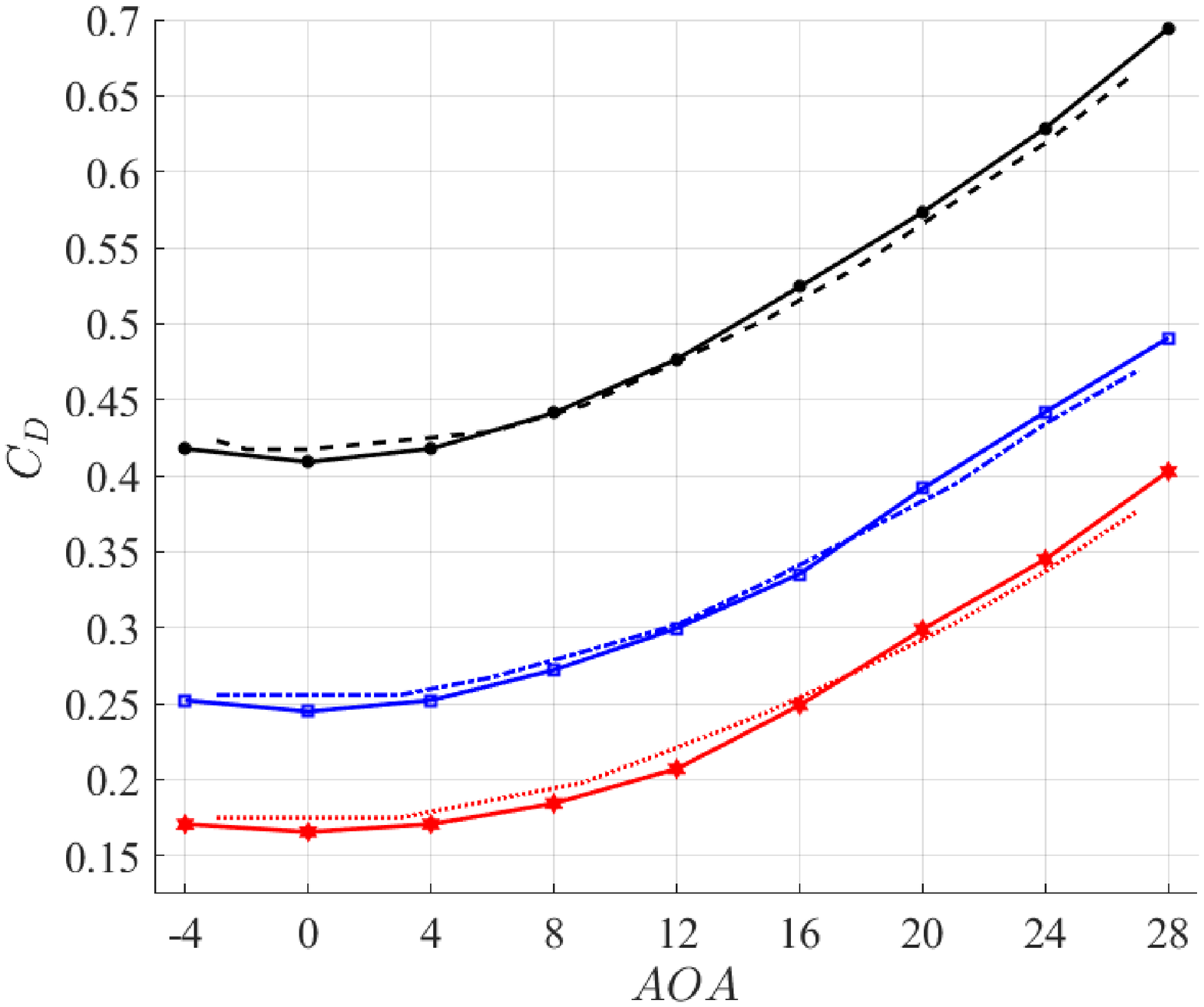}
		\caption{ Drag coefficient.}
		\label{fig:cd_AOA}
	\end{subfigure}
	\begin{subfigure}[H]{0.49\linewidth}
		\centering
		\includegraphics[width=\linewidth]{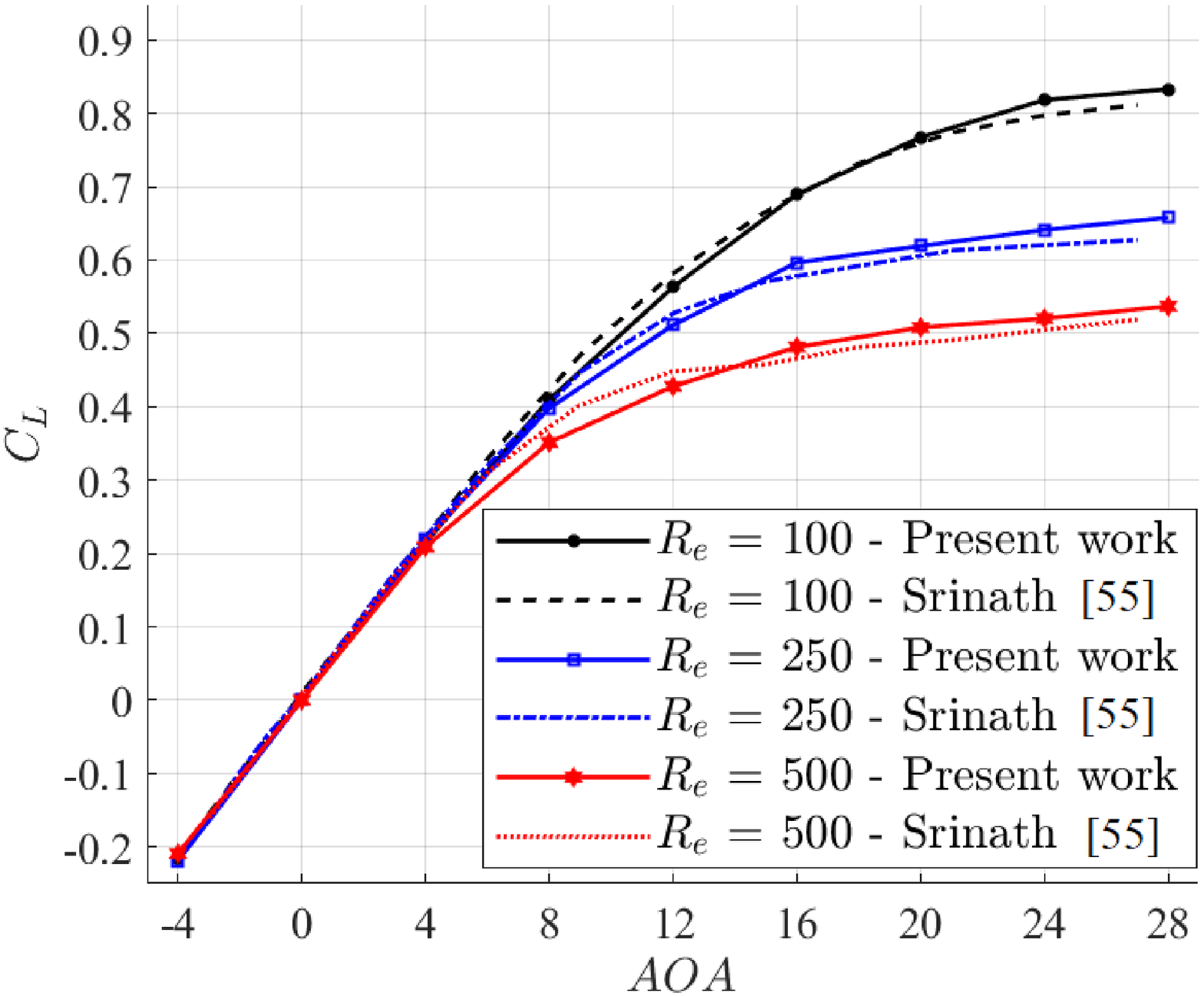}
		\caption{ Lift coefficient.}
		\label{fig:cl_AOA}
	\end{subfigure}
	\caption{Variation of hydrodynamic coefficients of a stationary NACA0012 foil with the angle of attack in an incompressible low-Reynolds number flow.}
    \label{fig:NACAFoilCLCD}
\end{figure}

The developed model's applicability to accurately determine the incompressible low-Reynolds number flow over a stationary NACA foil with different angles of attack is paramount, and therefore used as a validation case. The NACA0012 foil's location was obtained by implementing the LSM, see Figure \ref{fig:FlapFoil_Sta} (left), and the vorticity contours of the flow around the foils are shown in the right of Figure \ref{fig:FlapFoil_Sta}. Figure \ref{fig:NACAFoilCLCD} shows the validation of the numerical model by comparing the variation of lift coefficient and drag coefficients at different angles of attack with the numerical work of Srinath and Mittal \cite{srinath2009optimal}. Furthermore, the comparison shows the reliability and accuracy of the numerical capturing technique, LSM, as the maximum absolute error percentage of lift and drag coefficients is within an acceptable margin, $7\%$. 

\subsection{Oscillating Circular Cylinder}
\begin{figure}[H]
	\centering
	\includegraphics[width=\linewidth]{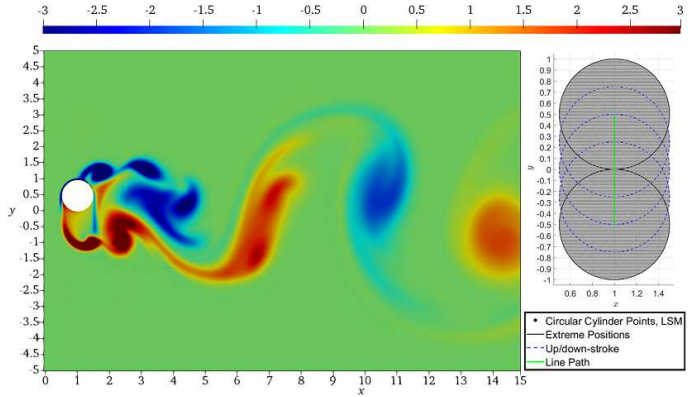}
	\caption{Vorticity contours around an oscillating circular cylinder at a Reynolds number of 185, the frequency ratio of 0.8, and amplitude ratio of 0.5 at the upper extreme position (left). Oscillating cylinder points at different time steps (right) using LSM.}
    \label{fig:StCyDiffRe2}
\end{figure}

\startsquarepar \indent Synchronization or lock-in, between the vortex shedding and vibration frequencies, is one of the main features of an oscillating circular cylinder, and the lock-in range becomes larger \stopsquarepar
\noindent as the amplitude of oscillation increases \cite{koopmann1967vortex}. Furthermore, this phenomenon appears in flapping foil turbines within a certain regime of the reduced velocity and the pivot point location \cite{wang2017structural}. To compare with Dekhatawala and Shah \cite{dekhatawala2019numerical}, the flow around a circulating cylinder was determined at $Re=185$ with a frequency ratio of 0.8 and amplitude ratio of 0.5. The location of the oscillating cylinder was obtained using the LSM, see Figure \ref{fig:StCyDiffRe2} (right). Dekhatawala and Shah \cite{dekhatawala2019numerical} obtained a root mean square of the lift coefficient of $0.608$ and a mean drag coefficient of $1.5557$, while the results of the present investigation are $0.6482$ and $1.2758$, respectively. The reasons for the deviation in the results could be the difference in the solver type used, the element type in the mesh, and the mesh quality near the cylinder. The vorticity contours for this case at the upper extreme position are shown in Figure \ref{fig:StCyDiffRe2} (left).
\subsection{Flapping-Foil with Simple Sinusoidal Motion}
\begin{figure}[H]
	\centering
	\includegraphics[width=0.85\linewidth]{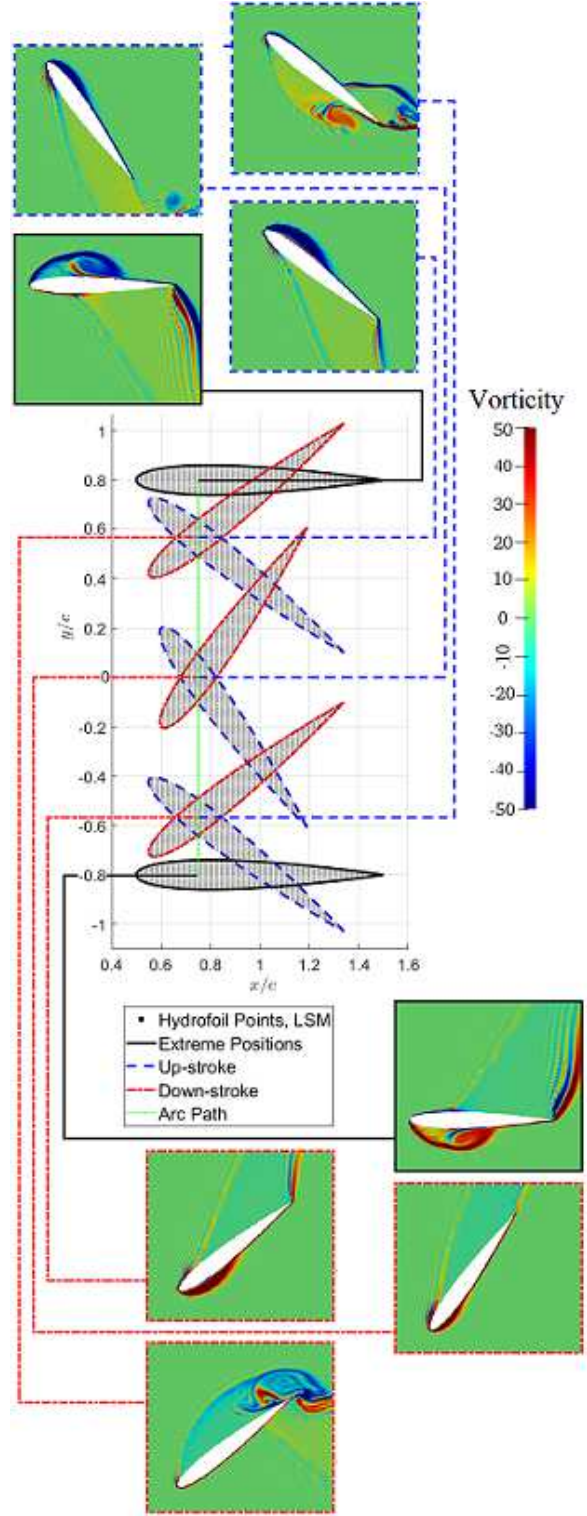}
	\caption{flapping foil (middle) and the Vorticity contours of the flow (right and left) at certain time steps using FEM and LSM at heave amplitude of $0.8$, pitch amplitude of $54^o$, Stroughal number of $0.25$, and Reynolds number of $1\times10^4$.}
	\label{fig:FlapFoil_Sim}
\end{figure}

\begin{figure}[H]
	\centering
	\includegraphics[width=\linewidth]{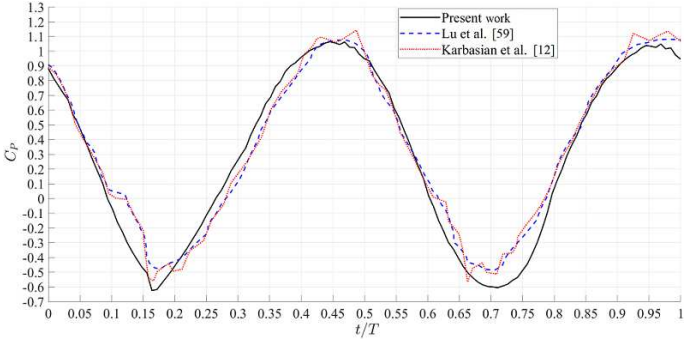}
	\caption{Power coefficient as a function of time for a NACA0012 flapping foil with simple sinusoidal motion at heave amplitude of $0.8$, pitch amplitude of $54^o$, Stroughal number of $0.25$, and Reynolds number of $1\times10^4$.}
	\label{fig:SimpleSinFlapFoil}
\end{figure}

Simple hydrokinetic turbines use sinusoidal flapping foils. LSM was used to capture the locations of the flapping foil, see the middle of Figure \ref{fig:FlapFoil_Sim}, and the vorticity contours of the flow around the flapping foil during the up- and down-stroke are shown in the right and left of the same figure, respectively.  Figure \ref{fig:SimpleSinFlapFoil} shows the power coefficient for a simple flapping NACA0012 foil at $Re=1\times10^4$, with a Stroughal number of $0.25$, heave amplitude of $0.8$, and pitch amplitude of $54^o$. Lu et al. \cite{lu2014nonsinusoidal} and Karbasian et al. \cite{karbasian2016power} obtained an efficiency (Equation \ref{eq:effi}) of $0.147$ and $0.17$, respectively. With the present solver, the efficiency was $0.216$. The deviation in results might be due to the difference in the selected numerical solvers' type between the present work (FEM solver) and their work (FVM solvers).
\section{Result and Discussion}
\begin{figure}[H]
	\centering
	\includegraphics[width=0.98\linewidth]{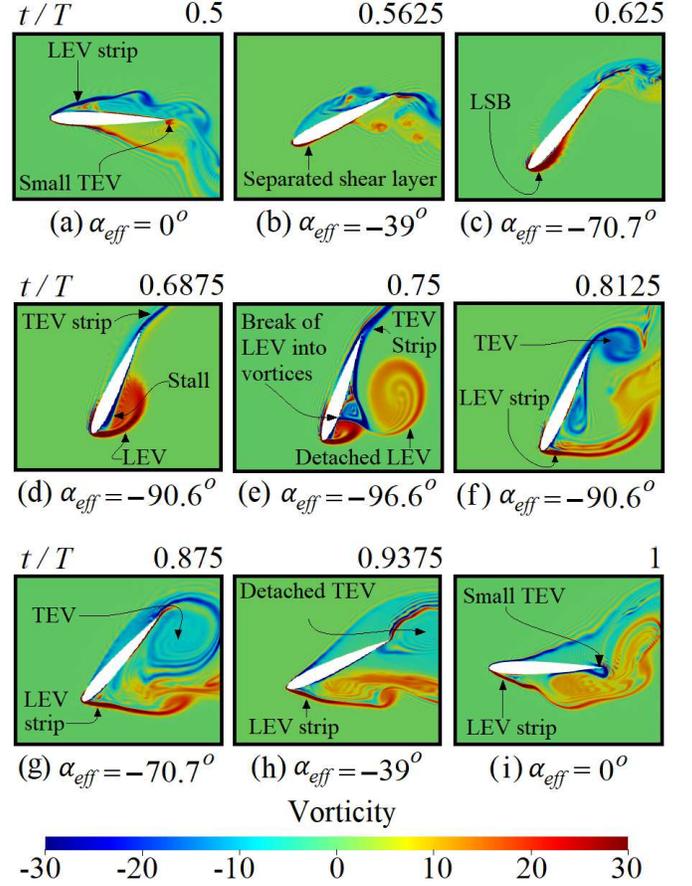}
	\caption{Vorticity contours around a NACA0012 flapping foil in swing-arm mode during the down-stroke phase at Reynolds number of $1\times10^5$, swing factor of $0.25$, and non-dimensional reduced frequency parameter of $0.08$.}
	\label{fig:VortexNACA0012}
\end{figure}

\begin{figure}[H]
	\centering
	 \begin{subfigure}[H]{\linewidth}
		 \centering
		\includegraphics[width=\linewidth]{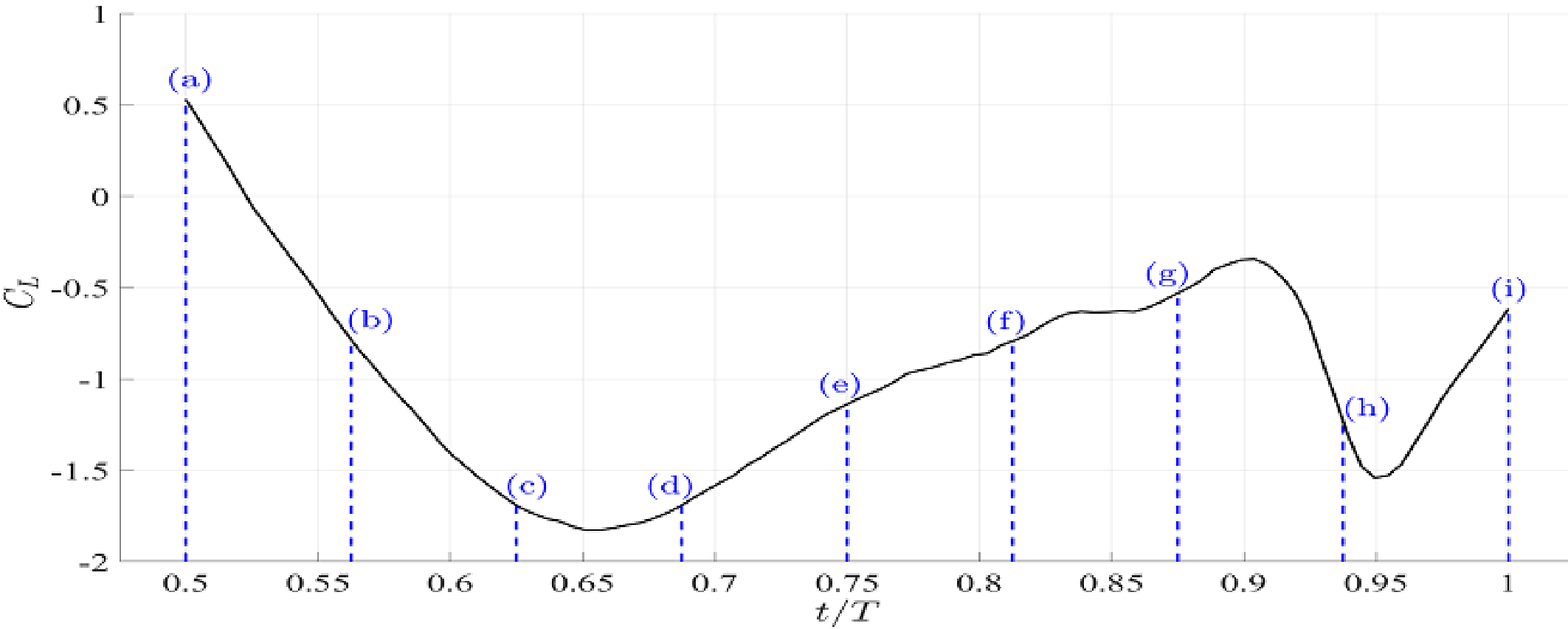}
		\caption{ Lift coefficient.}
		\label{fig:cl_NACA0012}
	\end{subfigure}\\
	\begin{subfigure}[H]{\linewidth}
		\centering
		\includegraphics[width=\linewidth]{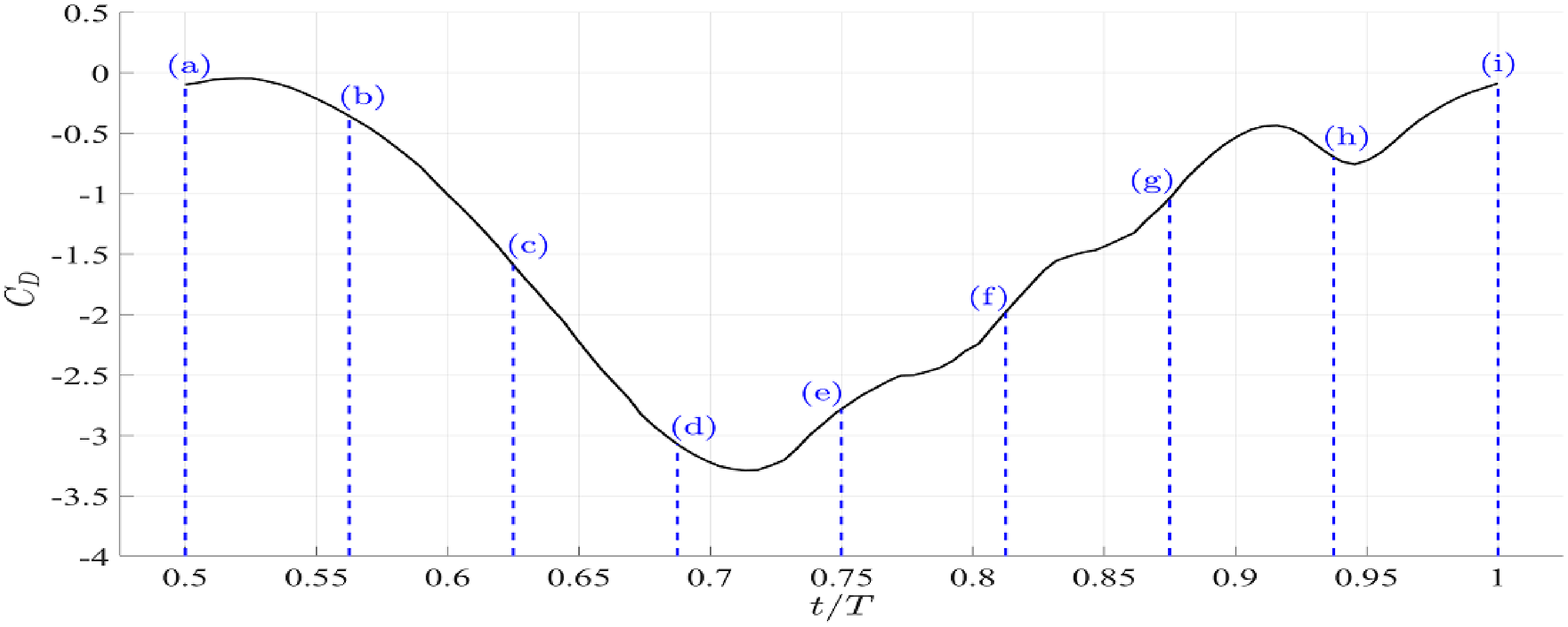}
		\caption{ Drag coefficient.}
		\label{fig:cd_NACA0012}
	\end{subfigure}\\
	\begin{subfigure}[H]{\linewidth}
		\centering
		\includegraphics[width=\linewidth]{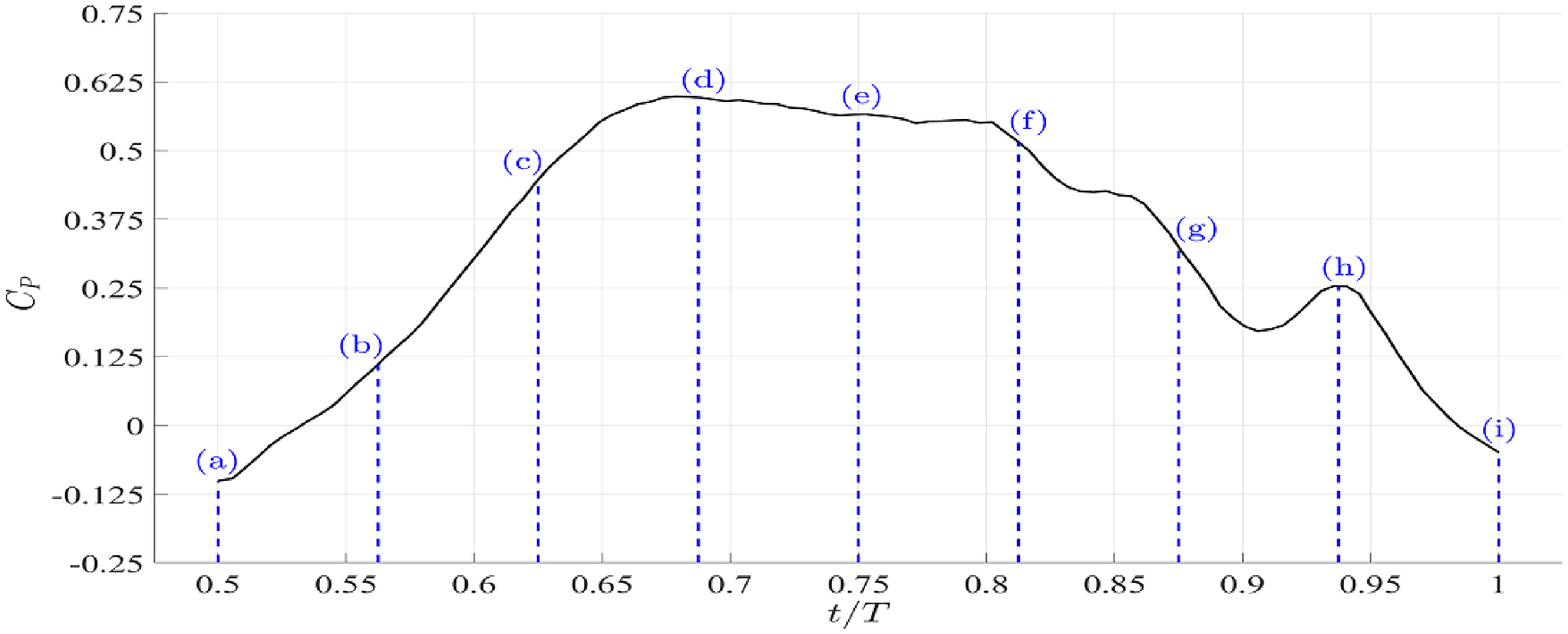}
		\caption{ Power coefficient.}
		\label{fig:cp_NACA0012}
	\end{subfigure}
	\caption{Lift, drag, and power coefficients of a NACA0012 flapping foil in swing-arm mode during the down-stroke phase at a Reynolds number of $1\times10^5$, swing factor of $0.25$, and non-dimensional reduced frequency parameter of $0.08$.}
    \label{fig:LDPNACA0012}
\end{figure}

In this study, a FEM model for the advancement of hydrokinetic flapping foil turbines was developed. To accurately predict the performance of flapping foil in the swing-arm correct capturing of the structure of the vortex around the foil is key. The flow around a NACA0012 foil during the down-stroke phase of its cycle is shown in Figure \ref{fig:VortexNACA0012}. The relation between the vortex formation, the lift, drag forces, and the extracted power are shown in Figure \ref{fig:LDPNACA0012}. Table \ref{table:1} shows the used parameters used in this study.

\begin{table}[H]
\centering
\caption{Parameters used in the numerical simulation.}
\begin{tabular}{c c c} 
 \hline
 Foil's chord & $c$ & 0.2 m\\ [0.5ex] 
 \hline
 Heave amplitude & $h_o$ & 1.25 m\\ 
 Pitch amplitude & $\theta_o$ & $70^o$\\
 Phase difference angle & $\psi$ & $90^o$\\
 Swing factor & $S$ & 0.25\\
 Non-dimensional reduced frequency & $k$ & 0.08\\
 Reynolds number & $Re$ & $1\times10^5$\\ [1ex] 
 \hline
\end{tabular}
\label{table:1}
\end{table}
At the start of the down-stroke, the foil has a zero effective angle of attack, Figure \ref{fig:VortexNACA0012} (a), and the power coefficient is at its lowest value, Figure \ref{fig:LDPNACA0012} (a). Here, there is a thin clockwise vortex strip at the leading-edge, which is created during the end of the up-stroke (the effective angle goes from positive to zero), and a small trailing-edge vortex (TEV) is developed, which reduces the lift force \cite{gharali2013dynamic}. Subsequently, as the effective angle of attack increases, an adverse pressure gradient appears. This leads to the creation of the separated shear layer, Figure \ref{fig:VortexNACA0012} (b), and a small increase in the power coefficient, see Figure \ref{fig:LDPNACA0012} (b). Figure \ref{fig:VortexNACA0012} (c) shows that the flow reversal layer grows with time, creating the Laminar Separation Bubble (LSB). The LSB is located at the leading-edge and is thus a small LEV. The LSB is the main reason for the increase in lift, drag forces and, as a result the extracted power as shown in Figure \ref{fig:LDPNACA0012} (c). As the LSB enlarges, the LEV forms. Figure \ref{fig:VortexNACA0012} (d) shows how the LEV travels towards the trailing edge during its growth. The stall phenomenon begins when the LEV is well-developed. During this phase, the flapping foil produces the highest possible lift and extracted power,  Figure \ref{fig:LDPNACA0012} (d). The break of LEV into small vortices causes detachment from the surface of the foil, and sequentially its strength decreases, as shown in Figure \ref{fig:VortexNACA0012} (e). This happens near the maximum value of the effective angle of attack. Figure \ref{fig:LDPNACA0012} (e) shows a slight decrease in the generated energy. As the value of the effective angle of attack continues to decrease, the TEV size quickly widens. This pushes the strip of LEV away from the foil's surface, as shown in Figures \ref{fig:VortexNACA0012} (f) and (g). Here, the lift, drag, and power coefficients significantly decrease. When the foil is close to finishing the down-stroke, the TEV moves away from the foil's surface and the LEV strip gets close the foil again, as shown in Figures \ref{fig:VortexNACA0012} (h) and (i). Here, a small bump in the extracted power occurs before reaching its lowest value again at a zero effective angle of attack (end of down-stroke), as shown in Figures \ref{fig:LDPNACA0012} (h) and (i).
\section{Conclusion}
\startsquarepar \indent The purpose of this study was to introduce a reliable numerical purpose-built code for the investigation of the hydrodynamics of hydrokinetic energy converters. The purpose-built solver is based on Finite Element code with the Level-Set technique. LSM is used to specify the foil nodes in the fixed grid, while FEM is applied to obtain the flow field around the flapping foil at each \stopsquarepar

\noindent time step. LSM reserves the quality of the grid while avoiding re-meshing that can lead to element inversion. The numerical model was validated using existing experimental and numerical results. Finally, a detailed vortex structure study of the flapping foil turbine in swing-arm mode is performed to show the relationship between the Leading-Edge Vortex and extracted power.

The main indication of the power extraction capability for the flapping foil is the strength of LEV. This is in line with the findings of Karbasian et al. \cite{karbasian2016power} and Xu et al. \cite{xu2019numerical}. Further, the gained power starts to fade out when the stall phenomenon appears. In other words, the flapping foil provides energy as long as the separation of LEV is delayed. Further numerical investigations, using FEM and LSM, to study the effect of the shape of the foil or the active-controlled high lift devices, such as flaps and slats, on power extraction from flapping foil hydrokinetic turbines in swing-arm mode is recommended.
\bibliography{Hamada}
\end{multicols*}
\end{document}